\documentclass[11pt,twoside]{article}
\usepackage{graphicx}
\usepackage{amsmath}
\usepackage{amssymb}
\usepackage{mathrsfs}

\begin{document}

\newcommand{\pst}{\hspace*{1.5em}}

\newcommand{\be}{\begin{equation}}
\newcommand{\ee}{\end{equation}}
\newcommand{\bm}{\boldmath}
\newcommand{\ds}{\displaystyle}
\newcommand{\bea}{\begin{eqnarray}}
\newcommand{\eea}{\end{eqnarray}}
\newcommand{\bdm}{\begin{displaymath}}
\newcommand{\edm}{\end{displaymath}}
\newcommand{\ba}{\begin{array}}
\newcommand{\ea}{\end{array}}
\newcommand{\arcsinh}{\mathop{\rm arcsinh}\nolimits}
\newcommand{\arctanh}{\mathop{\rm arctanh}\nolimits}
\newcommand{\bc}{\begin{center}}
\newcommand{\ec}{\end{center}}

\thispagestyle{plain}

\label{sh}

\begin{center} {\Large \bf
\begin{tabular}{c}
OPTICAL TOMOGRAPHY OF PHOTON-ADDED \\[-1mm]
COHERENT STATES, EVEN/ODD COHERENT STATES \\[-1mm]
AND THERMAL  STATES \\[-1mm]
\end{tabular}
 } \end{center}

\smallskip

\bigskip
\begin{center} {\bf
Ya. A. Korennoy, V. I. Man'ko
}\end{center}

\medskip

\begin{center}
{\it
P.N.    Lebedev Physics Institute,                          \\
       Leninsky prospect 53, 117924 Moscow, Russia
 }
\end{center}

\begin{abstract}\noindent
Explicit expressions for optical tomograms of the photon-added coherent
states, even/odd photon-added coherent states and photon-added thermal
states are given in terms of Hermite polynomials.
Suggestions for experimental homodyne detection of the
considered photon states are presented. 
\end{abstract}


\bigskip

%----------------------------------------------
\section{Introduction}
%----------------------------------------------
%P1************
\pst
Different kinds of nonclassical photon states were studied during last
two decades (see, e.g. reviews \cite{Dodonov2002}). Among the nonclassical
states there are so called photon-added and photon subtracted states
\cite{AgarwalTara}.

These states were considered in \cite{Kor2}, \cite{Kor3},
\cite{DodMizr}. In \cite{DodMizr1} it was shown that in photon subtracted states
the mean number of photons may be either  less or greater than of initial state
depending on the photon statistics, while in photon-added states the meal number
of photons is always greater then  that of initial state.   Experimentally these
states were studied in \cite{ZavattaBellini2004}, \cite{ZavattaBellini2007},
\cite{Paris}, \cite{Bellini3}, \cite{VogelBelliniZavatta}
 where the commutation relation of bosonic
annigilation and creation operators was considered. The experiments were done
by using the homodyne detection procedure of photon states applied
in \cite{Raymer93}, \cite{LvovscyRaymer} to reconstruct the Wigner function
of photon states.
%P2************
On the other hand the optical tomograms of the quantum state can be considered
as primary object which contain complete information of the quantum states and
can serve as alternative of density matrix \cite{Mancini96} , \cite{Ibort}.
In view of this the homodyne detection of photon states used in experiments
with photon-added states is interesting per se.
The homodyne detection provides as experimental output the optical tomogram
which is fair probability distribution of homodyne quadrature $X$ depending
on extra parameter $\theta$.
This angle parameter is so called local oscillator phase which is controlled
parameter varying in domain $0\geq\theta\geq2\pi$. Note, that this probability
satisfies the symmetry relation $w(X,\theta+\pi)=w(-X,\theta)$, and thus
in the range $0\geq\theta\geq\pi$ it completely describes the quantum state.

In \cite{ZavattaBellini2004} the authors used conditional stimulated
parametric down-conversion in a nonlinear optical crystal for the
production of the photon-added states. Here one high-energy pump 
photon can annihilate into two photons that obey the global energy and 
momentum conservation laws and are normally emitted into symmetrically 
oriented directions (signal and idler modes). To generate photon-added 
state one has to inject a seed (initial) state into a signal mode of 
the parametric amplifier. The conditional preparation of the target 
photon added state will be take place every time that a single photon 
is detected in the correlated idler mode.

The primary light source for the experiment is a mode-locked laser 
whose pulses are frequency-doubled to become the pump for degenerate 
parametric down-conversion in a type-I beta-barium borate crystal.

The Wigner functions of single photon added coherent states were 
reconstructed for experimentally measured data for 
$\vert\alpha\vert=0.03$, $0.9$, $2.6$.

In \cite{ ZavattaBellini2007} the authors have probed quantum 
commutation rules by addition and subtraction of single photons to/from a 
light field. They used the same procedure to generate photon-added, photon 
subtracted, photon-added-then subtracted,  and photon-subtracted-then-added 
states, which are different from each other and from the original seed 
thermal state.
They experimentally measured the quadrature distribution function of 
photon-added and photon-subtracted states for initial thermal state 
with mean number of photons $\bar n=0.57$.
Note, that this measured distribution function coincides with optical 
tomogram because the density matrix of this state exhibits no phase 
dependence. In development of this work in \cite{Bellini3} the authors 
measured tomograms of superpositions of photon-added-then-subtracted 
and photon-subtracted-then-added states , depending on phase between two
these  states.

In \cite{Paris} the parametric down-conversion in a nonlinear crystal 
also used to generate photon-added coherent states, and non-Gaussianity 
of these states were investigated. Experimental Wigner functions were 
reconstructed for single photon coherent states with $\vert\alpha\vert=0$,
$0.48$, $1.02$.

In \cite{VogelBelliniZavatta} Experimental preparation of single photon added 
thermal states and reconstruction of the Glauber-Sudarshan quasiprobability
($P$-function) for these states have been done.

It has been shown \cite{Kalamidas} that the higher order photon-added 
coherent states $\vert\alpha,m\rangle$ corresponding to $m > 1$, also can be 
realized in the parametric down-conversion scheme.

Generation of the photon-added coherent states also is possible in the 
cavity-atom interaction \cite{AgarwalTara}.

More over, the interaction in optomechanical systems (micro-resonator
interacting with laser light) by proper detuning can be used to generate
photon-added coherent states in the optomechanical domain \cite{Aspelmeyer}.

A common feature of the mentioned experimental techniques is that they are 
all used interacting bipartite systems. During evolution, the two subsystems 
are entangled. This enables to make suitable conditional measurements on one 
of the subsystems so that the other subsystem is prepared in a photon-added 
state.

In report \cite{Sivakumar} a comparative study of two processes, namely,
 parametric down-conversion and atom-cavity interaction, that can generate 
photon-added states is presented.
By expressing the time-evolved states in a suitable non-orthogonal
basis, it is established that the later method generates ideal photon-added 
states, a feature that is not present in the atom-cavity scheme. Further, the 
parametric down-conversion itself is shown to be capable of generating ideal 
$m$-photon-added state, without requiring higher order processes.

%P3*************
The aim of this work is to study optical tomograms of different kinds of
photon-added nonclassical states. We consider photon-added coherent states and
thermal states. Even/odd coherent states \cite{Dod74} with added photons
will be also studied in this paper.
The obtained below optical tomograms could be used to compare the
experimental results of the homodyne detection of the photon states, i.e. the
output experimental optical tomogram
(e.g. in  \cite{ZavattaBellini2004}, \cite{ZavattaBellini2007},
\cite{Paris}, \cite{Bellini3})  with the tomograms calculated in this
our paper. It is worth to note that some recent studies of photon-added
states are presented in \cite{Sivakumar}.

We also want to point out that the consideration of homodyne detection
of one-mode photon-added states in the tomographic probability
representation and the analysis of the experimental precision of the
obtained optical tomograms \cite{ZavattaBellini2004},
\cite{ZavattaBellini2007}, \cite{Paris}, \cite{Bellini3},
\cite{VogelBelliniZavatta} has the aspects which have to be discussed
in the connection with other experiments where different photon states
were studied, see, e.g. \cite{Zurich}.

In all the published results not all the tests related to properties
of the optical tomograms associated with checking the uncertainty relations
and studying the dependence of optical tomogram on local oscillator phase
have been given in detail and in explicit form. Another goal of our work
is to suggest to pay more attention to the variety of the possible tests
which permit to do better estimation of accuracy of the measuring such
primary object as optical tomogram identified with quantum state.

The paper is organized as follows In Sec. 2 short review of optical
tomographic representation of the photon states is given. In Sec.3 we review
a model of parametric oscillator and  photon-added states appeared in
this model are considered. The resume of results is presented in
conclusion  Sec.4

%P5
%---------------------------------
\section{Optical tomographic representation}
%---------------------------------
\pst

In this section we review the tomographic probability representation of
quantum states \cite {Ibort}. Given a density operator $\hat\rho$ (one mode
photon state). The symplectic tomogram $M(X,\mu,\nu)$ which is normalized
probability distribution of homodyne quadrature $X$ depending on two real
parameters $\mu$ and $\nu$ is defind
\be		\label{eq1}
M(X,\mu,\nu)=\mbox{Tr}\{\hat\rho\delta(X-\mu\hat q-\nu\hat p)\}
=\langle X,\mu,\nu\vert\hat\rho\vert X,\mu,\nu\rangle,
\ee
where $\hat q$ and $\hat p$ are operators of the first and second photon
quadratures, and $\vert X,\mu,\nu\rangle$ is an eigenvector of the hermitian
operator $\mu\hat q-\nu\hat p$ for the eigenvalue $X$.
For arbitrary state with wave function $\Psi(y)$ the tomogram is
expressed in terms of fractional Fourier transform of the wave function
\be		\label{eq2}
M(X,\mu,\nu)=\langle X,\mu,\nu\vert\Psi\rangle\langle\Psi\vert X,\mu,\nu\rangle
=\frac{1}{2\pi\vert\nu\vert}\left\vert\int\Psi(y)
~e^{\ds{\frac{i\mu}{2\nu}y^2-\frac{iXy}{\nu}}}\mbox{d}y\right\vert^2.
\ee

%P6

In experiments with homodyne detecting photon states the optical tomogram is
measured. This tomogram reads 
\be		\label{eq3}
w(X,\theta)= \mbox{Tr}\{\hat\rho\delta(X-\hat q\cos\theta-\hat p\sin\theta)\}
=\langle X,\theta\vert\hat\rho\vert X,\theta\rangle,
\ee
where $\vert X,\theta\rangle$ is an eigenvector of the hermitian
operator $\hat q\cos\theta-\hat p\sin\theta$ for the eigenvalue $X$.

The tomograms (\ref{eq1}) and(\ref{eq3}) are related
\be		\label{eq4}
w(X,\theta)= M(X,\cos\theta,\sin\theta)
\ee
\be		\label{eq5}
M(X,\mu,\nu)=\frac{1}{\sqrt{\mu^2+\nu^2}}~
w\left(\frac{X}{\sqrt{\mu^2+\nu^2}},\tan^{-1}\frac{\nu}{\mu}\right).
\ee
The density operator can be reconstructed as \cite{dAriano}
\be		\label{eq6}
\hat\rho=\frac{1}{2\pi}\int M(X,\mu,\nu)
~e^{i(X-\hat q\mu-\hat p\nu)}~\mbox{d}X~\mbox{d}\mu~\mbox{d}\nu.
\ee
In view of (\ref{eq5}) the density operators can be determined in terms of the
optical tomogram $w(X,\theta).$ The invertible relations of the measurable
tomograms and density operator provide possibility to use the optical tomogram
%P7
as primary object containing complete information of quantum states. In fact,
the quantum evolution equation and energy level equation were written for
optical tomogram in explicit form in \cite{Kor13}. The quadrature
statistics can be obtained from optical tomogram
\be		\label{eq7}
\langle X^n\rangle(\theta)=\int X^nw(X,\theta)~~\mbox{d}X.
\ee
Also the photon number statistics can be obtained from optical tomogram.
In fact,
\be		\label{eq8}
\langle\hat q^n \rangle=\int X^nw(X,\theta=0)~\mbox{d}X,
\ee
\be		\label{eq9}
\langle\hat p^n \rangle=\int X^nw(X,\theta=\pi/2)~\mbox{d}X.
\ee
Also mean photon number reads 
\be		\label{eq10}
\langle\hat n \rangle=\frac{1}{2}\int X^2\left[w(X,\theta=0)+
w(X,\theta=\pi/2)\right]\mbox{d}X~-\frac{1}{2}.
\ee

%P8
The experimental tomograms must satisfy uncertainty relations. These
relations in the tomographic form read
\bea		
&&\left[\int X^2w(X,\theta=0)~\mbox{d}X -\left(\int X~w(X,\theta=0)
~\mbox{d}X\right)^2\right] \nonumber \\[3mm]
&&~~~~~~~~~~\times 
\left[\int X^2w(X,\theta=\pi/2)~\mbox{d}X 
-\left(\int X~w(X,\theta=\pi/2)~\mbox{d}X\right)^2\right]
\geq\frac{1}{4}.
\label{eq11}
\eea
For example for coherent state the product in this relation equals $1/4$.

We point out that the optical tomography can provide possibility of
direct measuring the purity of photon quantum state \cite{Ventriglia}.

%----------------------------------------
\section{Optical tomography of photon-added states}
%-------------------------
\subsection{Photon-added coherent states}
\pst
Let us find tomograms of studied in \cite{Kor2} time dependent photon-added
coherent states $|\alpha,m,t\rangle$ of one mode parametric oscillator with
the Hamiltonian
\bdm
\hat H=\frac{\hat p^2}{2}+~\Omega^2(t)\frac{\hat q^2}{2},~~~~\Omega(0)=1.
\edm
The state $|\alpha,m,t\rangle$ is defined as follows:
\be				\label{Ua}
\vert\alpha,m,t\rangle=~\hat U(t)\vert\alpha,m,0\rangle= 
~\hat U(t)\vert\alpha,m\rangle=
\left(m!L_m(-|\alpha|^2)\right)^{-1/2}\hat U(t)
\hat a^{+m}\vert\alpha\rangle,
\ee
where $L_m(z)\equiv L_m^{(0)}(z)$ is the Laguerre polynomial
\cite{Bateman,Szego},
$\vert\alpha\rangle$ is the initial coherent state,
$\hat U(t)$ is the unitary
evolution operator
\be				\label{UU+}
~\hat U(t)~\hat U^+(t)=~\hat 1, 
~~~~~~\hat U(0)=~\hat 1.
\ee
As shown in \cite{Kor2}  the  expression for
the state  $\vert\alpha,m,t\rangle$ in the coordinate representation
has the form
\be
\langle q\vert\alpha,m,t\rangle=
\left(m!L_m(-|\alpha|^2)\right)^{-1/2}
\left(\ds{\frac{\varepsilon^*}{2\varepsilon}}\right)^{m/2}
~H_m\left(\ds{\frac{q}{\vert\varepsilon\vert}-\sqrt{\frac{\varepsilon^*}
{2\varepsilon}}\alpha}\right)\langle q\vert\alpha,t\rangle,
\ee
where $H_m(z)$ is the Hermite polynomial \cite{Bateman,Szego},
$\langle q\vert\alpha,t\rangle$
is the time-dependent coherent state
\be				\label{timeCohSt}
\langle q\vert\alpha,t\rangle =
\pi^{-1/4}\varepsilon ^{-1/2}
\exp\left(\frac{i\dot\varepsilon q^2}{2\varepsilon}+
\frac{\sqrt 2\alpha q}{\varepsilon}
-\frac{\alpha^2\varepsilon^*}{2\varepsilon}
-\frac{\vert\alpha\vert^2}{2}\right), 
\ee
which was considered, e.g. in \cite{MM70},
 $c$-number function $\varepsilon(t)$ satisfies the equation
\be			\label{e-equation}
~\ddot\varepsilon(t)+~\Omega^2(t)~\varepsilon(t)=0,
\ee
with the initial conditions $~\varepsilon(0)=1, ~~~\dot\varepsilon(0)=i,~~$
which means that the Wronskian is
\be				\label{Vronscian}
\varepsilon\dot\varepsilon^*-
\varepsilon^*\dot\varepsilon=-2i.
\ee
The state $|\alpha,m,t\rangle$ is pure, thus we can find tomogram of
it from the formular (\ref{eq2}).
After some calculations we get
\begin{eqnarray}
M_{\alpha m}(X,\mu,\nu,t)&=&\frac{\left(m!L_m(-|\alpha|^2)\right)^{-1}}
{\sqrt\pi2^m
|\dot\varepsilon\nu+\varepsilon\mu|}
\left|H_m\left\{\left(\frac{X\varepsilon+i\sqrt2\alpha\nu}
{|\varepsilon|(\mu\varepsilon+\nu\dot\varepsilon)}
-\sqrt{\frac{\varepsilon^*}{2\varepsilon}}\alpha\right)
\left(\frac{|\varepsilon|^2(\mu\varepsilon+\nu\dot\varepsilon)}
{\varepsilon^2(\mu\varepsilon^*+\nu\dot\varepsilon^*)}\right)^{1/2}
\right\}\right|^2 \nonumber \\[3mm]
{}&\times&\left|\exp\left\{-\frac{|\alpha|^2}{2}
-\frac{X^2}{2|\mu\varepsilon+\nu\dot\varepsilon|^2}
+\frac{\sqrt2\alpha X}{\mu\varepsilon+\nu\dot\varepsilon}
-\frac{\alpha^2\varepsilon^*}{2\varepsilon}
+\frac{i\nu\alpha^2}{\varepsilon(\mu\varepsilon+\nu\dot\varepsilon)}
\right\}\right|^2.
\label{eq_62}
\end{eqnarray}
The substitutions $\mu=\cos\theta$ and $\nu=\sin\theta$ to (\ref{eq_62})
gives us according to (\ref{eq4}) the optical tomogram $w_{\alpha m}
(X,\theta,t).$ 
In the case of stationary Hamiltonian ($\Omega(t)=1$, $\varepsilon=e^{it}$)
we have
\begin{eqnarray}
w_{\alpha m}(X,\theta,t,\Omega=1)&=&
\frac{\left(m!L_m(-|\alpha|^2)\right)^{-1}}{\sqrt\pi2^m}
\left|H_m\left(X-\frac{\alpha}{\sqrt{2}}e^{-i(t+\theta)}
\right)\right|^2 \nonumber \\[3mm]
&\times&\exp\left\{-X^2-\vert\alpha\vert^2+2\sqrt2X\mbox{Re}
(\alpha e^{-i(t+\theta)})-\mbox{Re}(\alpha^2e^{-2i(t+\theta)})\right\}
\end{eqnarray}
%%%%%%%%%%%%%%%%%%%%%%%%%%%%%%%%%%%%%
%%%%%%%%%%%%%%%%%%%%%%%%%%%%%%%%%%%%%
\begin{figure}[h]
\begin{minipage}[h]{0.49\linewidth}
\center{\includegraphics[width=1\linewidth,height=1\linewidth]
{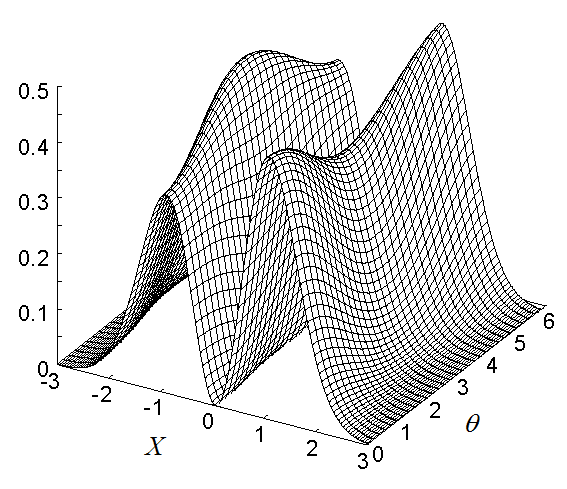} \\ a)}
\end{minipage}
%\hfill
\begin{minipage}[h]{0.49\linewidth}
\center{\includegraphics[width=1\linewidth,height=1\linewidth]
{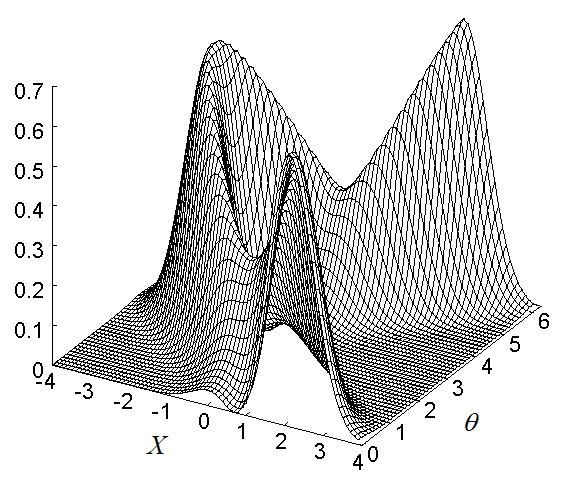} \\ b)}
\end{minipage}
\caption{\label{figure1} Optical tomograms $w_\alpha(X,\theta)$
 of photon-added coherent states for $\Omega(t)=1$,
$m=1$, $\alpha=0.1$ (a), $\alpha=1$ (b). }
\end{figure}
%%%%%%%%%%%%%%%%%%%%%%%%%%%%%%%%%%%%%
%%%%%%%%%%%%%%%%%%%%%%%%%%%%%%%%%%%%%

%--------------------------------------
\subsection{Photon-added even/odd coherent states}
%--------------------------------------
\pst
Photon-added even/odd coherent states $\vert\alpha_\pm, m\rangle$
were introduced in \cite{Kor3} as follows
\be		\label{QSOe1}
\vert\alpha_\pm, m\rangle=\frac{\hat a^{\dag m}\vert\alpha_\pm\rangle}
{(\langle\alpha_\pm\vert\hat a^m \hat a^{\dag m}\vert\alpha_\pm\rangle)^{1/2}},
\ee
where $\vert\alpha_\pm\rangle$ is an even/odd coherent states \cite{Dod74},
\cite{Ansari} and $m$ is an integer. The normalized states
$\vert\alpha_\pm\rangle$ can be written as
\bdm
\vert\alpha_+\rangle=\frac{e^{\vert\alpha\vert^2/2}}
{2\sqrt{\cosh{\vert\alpha\vert^2}}}
(\vert\alpha\rangle~+~\vert -\alpha\rangle),
\edm
\bdm
\vert\alpha_-\rangle=\frac{e^{\vert\alpha\vert^2/2}}
{2\sqrt{\sinh{\vert\alpha\vert^2}}}
(\vert\alpha\rangle~-~ \vert -\alpha\rangle).
\edm
Photon-added even/odd coherent states $\vert\alpha_\pm, m\rangle$ in terms
of photon-added coherent states $\vert\alpha, m\rangle$ are represented as
follows
\cite{Kor3}:
\be		\label{q1}
|\alpha_\pm,m\rangle=\left(\frac{e^{|\alpha|^2}L_m(-|\alpha|^2)}
{2(e^{|\alpha|^2}L_m(-|\alpha|^2)\pm
e^{-|\alpha|^2}L_m(|\alpha|^2))}\right)^{1/2}
~(|\alpha,m\rangle~\pm~|-\alpha,m\rangle)
\ee
This formula in view of (\ref{eq1}) and (\ref{Ua}) enables us to get the result
\begin{eqnarray}
M_{\alpha_\pm m}(X,\mu,\nu,t)&=&\frac{e^{|\alpha|^2}L_m(-|\alpha|^2)}
{2(e^{|\alpha|^2}L_m(-|\alpha|^2)\pm
e^{-|\alpha|^2}L_m(|\alpha|^2))} \nonumber \\[3mm]
&\times&\left\{
M_{\alpha m}(X,\mu,\nu,t)~+~M_{-\alpha m}(X,\mu,\nu,t) \right. \nonumber \\[3mm]
&\pm& \left.2\mbox{Re}~[\langle X,\mu,\nu|\alpha,m,t\rangle 
\langle-\alpha,m,t|X,\mu,\nu\rangle]\right\}, \label{q2}
\end{eqnarray}
and with respect to (\ref{eq4}) we have
$w_{\alpha_\pm m}(X,\theta,t)=M_{\alpha_\pm m}(X,\mu=\cos\theta,
\nu=\sin\theta,t)$,
and in the case of stationary Hamiltonian with $\Omega(t)=1$ we have
 $w_{\alpha_\pm m}(X,\theta,t)=w_{\alpha_\pm m}(X,t+\theta)$.

%%%%%%%%%%%%%%%%%%%%%%%%%%%%%%%%%%%%%
%%%%%%%%%%%%%%%%%%%%%%%%%%%%%%%%%%%%%
\begin{figure}[h!]
\begin{minipage}[h]{0.49\linewidth}
\center{\includegraphics[width=1\linewidth,height=1\linewidth]
{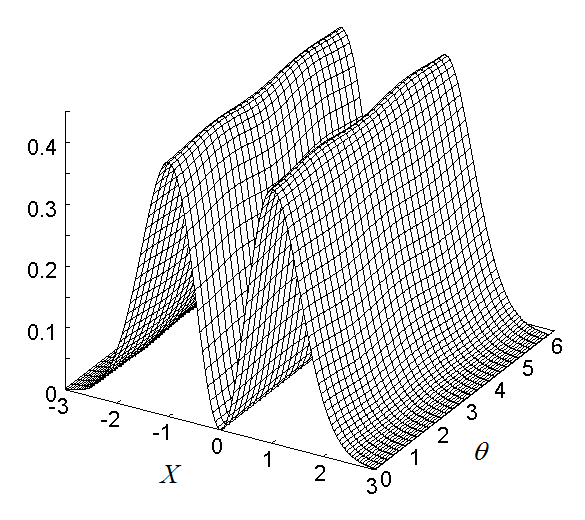} \\ a)}
\end{minipage}
%\hfill
\begin{minipage}[h]{0.49\linewidth}
\center{\includegraphics[width=1\linewidth,height=1\linewidth]
{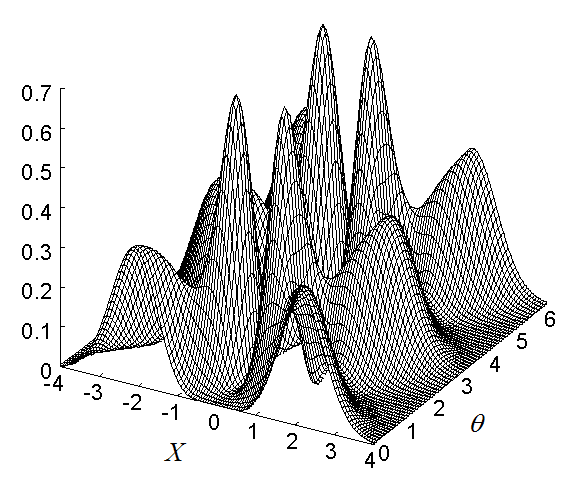} \\ b)}
\end{minipage}
\caption{\label{figure2} Optical tomograms $w_{\alpha_+}(X,\theta)$
of even photon-added coherent states
for $\Omega(t)=1$, $m=1$, $\alpha=0.1$ (a), $\alpha=1$ (b).}
\end{figure}
%%%%%%%%%%%%%%%%%%%%%%%%%%%%%%%%%%%%%
%%%%%%%%%%%%%%%%%%%%%%%%%%%%%%%%%%%%%
%%%%%%%%%%%%%%%%%%%%%%%%%%%%%%%%%%%%%
%%%%%%%%%%%%%%%%%%%%%%%%%%%%%%%%%%%%%
\begin{figure}[h!]
\begin{minipage}[h]{0.49\linewidth}
\center{\includegraphics[width=1\linewidth,height=1\linewidth]
{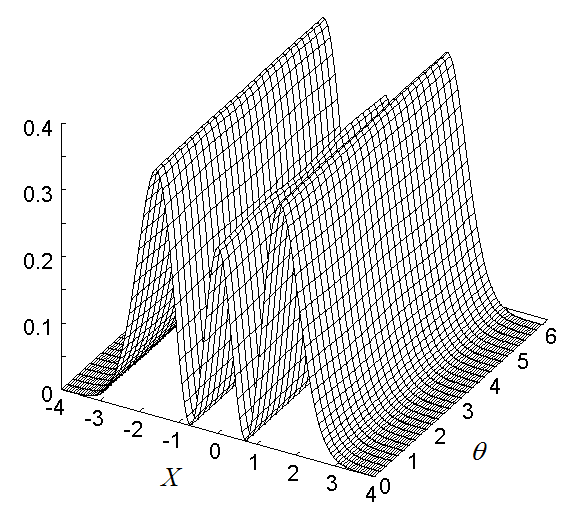} \\ a)}
\end{minipage}
%\hfill
\begin{minipage}[h]{0.49\linewidth}
\center{\includegraphics[width=1\linewidth,height=1\linewidth]
{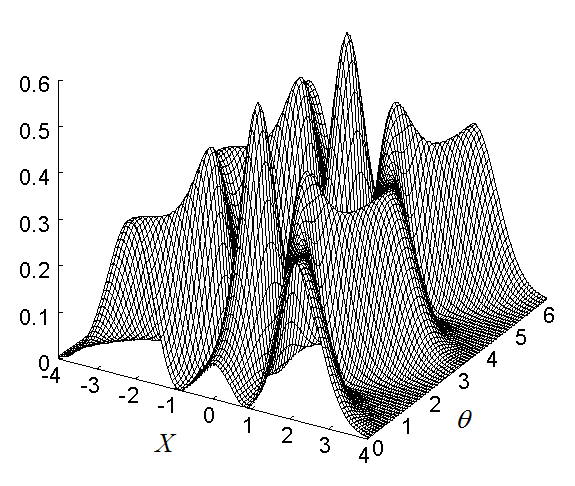} \\ b)}
\end{minipage}
\caption{\label{figure3} Optical tomograms $w_{\alpha_-}(X,\theta)$ 
of odd photon-added coherent states
for $\Omega(t)=1$, $m=1$, $\alpha=0.1$ (a), $\alpha=1$ (b).}
\end{figure}
%%%%%%%%%%%%%%%%%%%%%%%%%%%%%%%%%%%%%
%%%%%%%%%%%%%%%%%%%%%%%%%%%%%%%%%%%%%

%--------------------------------------
\subsection{Photon-added thermal states}
\begin{figure}[h!]
\begin{minipage}[h]{0.49\linewidth}
\center{\includegraphics[width=1\linewidth,height=1\linewidth]
{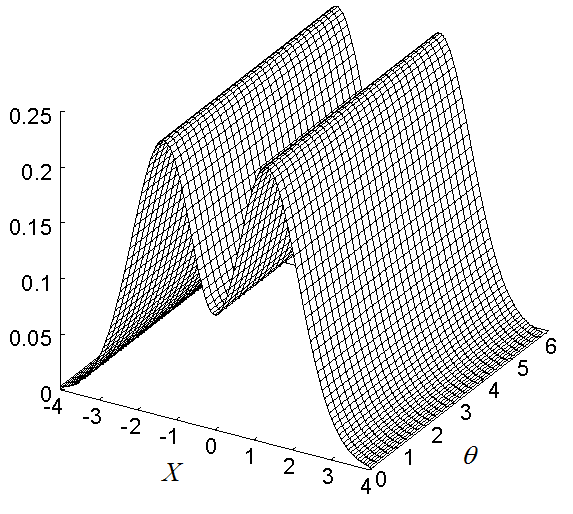} \\ a)}
\end{minipage}
%\hfill
\begin{minipage}[h!]{0.49\linewidth}
\center{\includegraphics[width=1\linewidth,height=1\linewidth]
{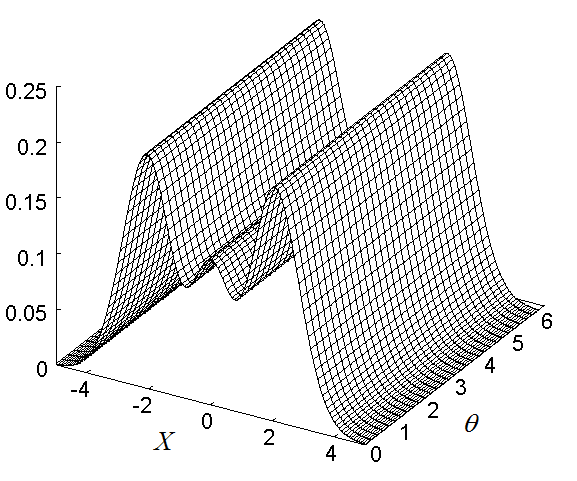} \\ b)}
\end{minipage}
\caption{\label{figure4} Optical tomograms $w_{Tm}(X,\theta)$ 
of thermal  states for $T=1$, $m=1$ (a),
$m=2$ (b).}
\end{figure}

%--------------------------------------
\pst
A thermal state is the most classical state of light, formed by a statistical
mixture of coherent states. The density operator of thermal state has the form
$\hat{\rho}_{\rm T} = Z^{-1} e^{-\hat{H}/T}$, $Z={\rm
Tr}e^{-\hat{H}/T}$. The associated optical tomogram reads $w_{\rm
T}(X,\theta) =\ds{ \frac{1}{\sqrt{2\pi\sigma^2}}e^{-X^2/2\sigma^2}}$,
where $\sigma^2 = \frac{1}{2}\coth\frac{1}{2T}$.
The density matrix of the photon-added thermal states in  representation 
of the Fock states $\vert n,t\rangle$ is given by
\be		\label{q3}
\langle k,t\vert\hat\rho_{Tm}\vert n,t\rangle
=\langle k,t\vert\hat a^{\dag m}\hat\rho_T~\hat a^m\vert n,t\rangle 
=\delta_{kn}\frac{(1-e^{-1/T})^{m+1}}{m!}
\frac{n!}{(n-m)!}e^{-(n-m)/T},~~~~(k,~~n\geq m).
\ee
The symplectic tomogram of this state  can be written as follows
\be		\label{q4}
M_{Tm}(X,\mu,\nu,t)=\sum_{n=0}^\infty\langle n+m,t\vert\hat\rho_{Tm}
\vert n+m,t\rangle
\vert\langle X,\mu,\nu\vert n+m,t\rangle\vert^2.
\ee
Noting that $\vert\langle X,\mu,\nu\vert n,t\rangle\vert^2=M_{\alpha=0,n}
(X,\mu,\nu,t)$,
with the help of (\ref{eq_62})  we can find the result
\bea		
M_{Tm}(X,\mu,\nu,t)&=&\frac{(1-e^{-1/T})^{m+1}}{\sqrt\pi~m!~2^m
~\vert\mu\varepsilon+\nu\dot\varepsilon\vert}
~e^{-X^2/\vert\mu\varepsilon+\nu\dot\varepsilon\vert^2} \nonumber \\[3mm]
&\times&\sum_{n=0}^\infty\frac{e^{-n/T}}{n!~2^n}
~\left\vert H_{n+m}\left\{\frac{X\varepsilon}
{|\varepsilon|(\mu\varepsilon+\nu\dot\varepsilon)}
\left(\frac{|\varepsilon|^2(\mu\varepsilon+\nu\dot\varepsilon)}
{\varepsilon^2(\mu\varepsilon^*+\nu\dot\varepsilon^*)}\right)^{1/2}
\right\}\right\vert^2.
\label{qHermite0}
\eea
For the stationary Hamiltonian the optical tomogram can be read
\be		\label{qHermite}
w_{Tm}(X,\theta)=\frac{(1-e^{-1/T})^{m+1}}{\sqrt\pi~m!~2^m}~e^{-X^2}
\sum_{n=0}^\infty\frac{e^{-n/T}H_{n+m}^2(X)}{n!~2^n}.
\ee
Thus the optical tomogram  of the photon-added thermal  state for 
the stationary Hamiltonian is not depend
on time and on the parameter $\theta$.
This formula with the help of 
the known summing expression  for Hermite polynomials
\cite{Bateman}
\bdm
(1-\eta^2)^{-1/2}\exp\left(\frac{2xy\eta-(x^2+y^2)\eta^2}{1-\eta^2}\right)
=\sum_{n=0}^\infty \frac{H_n(x)H_n(y)}{2^n~ n!}\eta^n,
\edm
is transformed to the differential expression
for  $w_{Tm}(X,\theta)=w_{Tm}(X)$
\be		\label{q5}
w_{Tm}(X)=\frac{(1-e^{-1/T})^{m+1}}{\sqrt{\pi}~m!}e^{-X^2}
\frac{\partial^m}{\partial\eta^m}
\left(\frac{\exp\left(\frac{2X^2\eta}{1+\eta}\right)}{\sqrt{1-\eta^2}} 
\right)_{\eta=e^{-1/T}}.
\ee
It gives us for $m=1$ and $m=2$ the following relations:
\be		\label{qm1}
w_{Tm=1}(X)=\frac{(1-e^{-1/T})^2}{\sqrt{\pi}\sqrt{1-e^{-2/T}}}
~e^{-X^2\tanh{(1/2T)}}
\left\{\frac{2X^2}{(1+e^{-1/T})^2}+\frac{e^{-1/T}}{1-e^{-2/T}}\right\};
\ee
\bea		
w_{Tm=2}(X)&=&\frac{(1-e^{-1/T})^3}{2\sqrt{\pi}\sqrt{1-e^{-2/T}}}
~e^{-X^2\tanh{(1/2T)}} \nonumber \\[3mm]
&\times&\left\{\frac{4X^4}{(1+e^{-1/T})^4}+
\frac{4X^2(2e^{-1/T}-1)}{(1+e^{-1/T})^2(1-e^{-2/T})}+
\frac{2e^{-2/T}+1}{(1-e^{-2/T})^2}\right\}.
\label{qm2}
\eea
In Figures \ref{figure1} - \ref{figure4} we illustrated tomograms 
of photon-added  states for some examples.
%%%%%%%%%%%%%%%%%%%%%%%%%%%%%%%%%%%%%
%%%%%%%%%%%%%%%%%%%%%%%%%%%%%%%%%%%%%
%%%%%%%%%%%%%%%%%%%%%%%%%%%%%%%%%%%%%
%%%%%%%%%%%%%%%%%%%%%%%%%%%%%%%%%%%%%

%----------------------------
\section{Conclusion}
%----------------------------
\pst
To resume we point out the main results of our paper. We calculated the
optical tomograms of photon-added states. The initial states to which the
photon are added are chosen as coherent state, thermal  state, even and
odd coherent state. We considered these states since they have been realized
experimentally in \cite{ZavattaBellini2004},\cite{ZavattaBellini2007},
\cite{Bellini3} . We pointed out that the optical tomograms
in probability representation of quantum mechanics can be considered as
primary objects containing complete information on quantum states
(see, e. g. \cite{Ibort}).

In view of this the reconstructing the Wigner function or Husimi function do
not add extra information on the state property. Thus the problem is to
measure the optical tomogram with highest possible accuracy. The precision of
the experiments for measuring optical tomogram can be checked using such
criteria as quadrature uncertainty relation with purity (temperature)
dependent bound \cite{DodonovMankoT183}. This criteria was suggested in
\cite{Ventriglia} to control precision of the homodyne experiments. In the
mentioned experimental works \cite{ZavattaBellini2004},
\cite{ZavattaBellini2007}, \cite{Paris}, \cite{Bellini3},
\cite{VogelBelliniZavatta},
including the homodyne two-mode
studies \cite{Zurich} the checking correspondence to the above criteria has not
been done. It is especially important, taking into account that the local 
oscillator
phase contribution can influence the final results if the states under the study
differ from the states (thermal or photon-added thermal) with optical
tomograms not depending on the local oscillator phase. 

It is of special interest to test the experimental optical tomogram in
the sense that it must satisfy the discussed in introduction symmetry
condition with respect to shift of local oscillator phase by $\pi$.
In view of this we suggest to measure accurately the optical tomograms
not only in domain of local oscillator phase $0\geq\theta\geq\pi$ but
to investigate accurately also the domain $\pi\geq\theta\geq2\pi$ to
check the symmetry properties of the tomogram.
Using the theoretically calculated optical homodyne probabilities as a 
standard for comparison with the experimental results, one can
estimate measure of necessity of  the precision improvement of the 
technique in future experiments.

\section{Acknoledgments}V.I.M. thank the Russian Foundation for Basic
Research for partial support under Projects Nos. 09-02-00142 and 10-02-00312.

\end{document}